\documentclass[a4paper]{jpconf}
\usepackage{graphicx}
\usepackage{amsmath}	
\usepackage{amssymb}	

\def\be{\begin{equation}}
\def\ee{\end{equation}}
\begin{document}
\title{Effects of magnetic fields in white dwarfs}

\author{B. Franzon}
\address{FIAS, Ruth-Moufang 1, 60438 Frankfurt am Main, Germany}
\ead{franzon@fias.uni-frankfurt.de}

\author{S. Schramm}
\address{FIAS, Ruth-Moufang 1, 60438 Frankfurt am Main, Germany}

\ead{schramm@fias.uni-frankfurt.de}

\begin{abstract}
We perfom calculations of white dwarfs endowed with strong magnetic fields. White dwarfs are the progenitors of supernova Type Ia explosions and they are widely used as candles to show that the Universe is expanding and accelerating. However, observations of ultraluminous supernovae have suggested that the progenitor of such an explosion should be a white dwarf with mass above the well-known Chandrasekhar limit $\sim \,$1.4 $\rm{M_{\odot}}$. In corroboration with other works, but by using a fully general relativistic framework, we obtained also  strongly magnetized white dwarfs  with masses $\rm{M  \sim 2.0\,\, M_{\odot}}$. 
\end{abstract}

\section{Introduction}
\hfill \break

 White dwarfs (WD) are formed in catastrophic astrophysical events such as supernova explosions. It is also in these events that other forms of compac objects, as neutron stars or black holes, can be created. It is  understood that stars born with masses less than $\sim$ 8
solar masses end up their evolutions as white dwarfs, see e.g.
Ref.~\cite{shapiro2008black}. With a typical composition  mostly made of
carbon, oxygen, or helium, endowed with a background of electron-degenerate matter,  white dwarfs are extremely dense objects, with a mass comparable to that of the Sun, which is distributed in a volume comparable to that of the Earth. As a result, the densities in WD's can reach values up to 
$\mathsf{\sim 10^{11}\, g/cm^{3}}$ at the stellar center. 

The mass of white dwarfs is primarily given by the total mass of the nuclei, whereas the main contribution to the pressure comes from the electrons. In addition, in Ref.~\cite{chandrasekhar1931maximum} was  found that there is a limit in the stellar mass, beyond which degenerate white dwarfs are unstable. This critical mass is the so-called  Chandrasekhar limit and is about 1.4 $\sf{M_{\odot}}$. 
 

 
 White dwarfs are not only very dense objects, but they are associated also with fast rotation (see Ref.~\cite{boshkayev2013general} and references therein)
  and also strongly magnetized.  Observations have shown that the observed surface magnetic fields in white dwafs can be so strong as $\mathsf{10^{9}}\,$ G \cite{Terada:2007br,Reimers:1995ia}.  Although the internal magnetic fields of these stars cannot be directly constrained by observations, according to Virial theorem arguments,  which give an upper estimate for the magnetic inside neutron stars, they can possess central magnetic fields as large as $\mathsf{10^{13}}$ G. This value is obtained in Newtonian mechanics by
equating the  stellar magnetic energy stored within the star, $\mathsf{B^2/8\pi \times 4\pi/ 3R^3}$, and  the gravitaional binding energy, $\mathsf{3GM^2/5R}$. In doing do, 
the magnetic field strength satisfies $\mathsf{B \sim 1.4\times 10^{14}\,(1.40\, M_{\odot}/M) (R/\,1000\, km)^{-2}}$ G.  On the other hand, analytic and numeric calculations, both in Newtonian theory as well as in General Relativity theory, showed that WD's may
have internal magnetic fields as large as $\mathsf{10^{12-16}}\,$ G \cite{shapiro2008black,
 Bera:2014wja,Franzon:2015gda}.
 
The influence of magnetic fields on the structure of white dwarfs is an important problem, since super-massive magnetized WD's, whose existence is partially supported by magnetic forces, could simplify the explanation of observed ultra-luminous explosions of supernovae Type Ia (SNe Ia). SNe Ia  occur  in  binary  systems,  when  a white  dwarf  accreting  matter from  a companion  approaches  the Chandrasekhar mass limit of 1.40 $\mathsf{M_{\odot}}$ and becomes unstable. As a result,  thermonuclear explosion  develops and an immense   amount of energy is suddenly released. A feature of these events is that SNe Ia have spectra and  light  curves (evolution of the supernova brightness with time) amazingly  uniform,  indicating  a  common  origin  and  a  common  intrinsic  luminosity.  However, recent observations of several supernovae appear to be more luminous than expected. This would indicate that such explositions occur when the white dwarf has a mass well above the Chandrasekhar mass limit.  

Previous calculations showed that magnetic white dwarfs can have their masses increased up to 2.58 $\sf{M_{\odot}}$ for a magnetic field strength of $\mathsf{10^{18}}\,$G at the center of the star \cite{Das:2013gd}. Nonetheless, such approach violates not only macro physics aspects, as for example, the breaking of spherical symmetry due to the magnetic field, but also microphysics considerations, which are relevant for a self-consistent calculation of the structure of these objects \cite{chamel2013stability}. In addition, a self-consistent newtonian structure calculation of strongly magnetized white dwarfs showed that these stars exceed the traditional Chandrasekhar mass limit significantly $(\sf{M \sim 1.9\, M_{\odot}} )$ for a maximum field strength of the order of $\sf{10^{14}\,}$G  \cite{Bera:2014wja}.

Here, we model static and rotating magnetized white dwarfs in a self-consistent way  by solving the coupled Einstein-Maxwell equations  by means of a pseudo-spectra method. The presence of such a strong magnetic field can locally affect the microphysics  of the stellar matter, as for example, due to Landau quantization.  However, in Ref.~\cite{Bera:2014wja} was shown that  Landau quantization does not affect the global properties of white dwarfs. Globally, the magnetic field can affect the structure of WD's, since it contributes to the Lorenz force, which acts against gravity. In addition, it contributes also to the structure of spacetime, since the magnetic field is now a source for the gravitational field through the Maxwell energy-momentum tensor. In the following, as we are interested in global effects that magnetic fields and rotation can induce in WD's, we simplify the discussion assuming white dwarfs that are predominately composed by $\mathsf{^{12}C}$ ($\mathsf{A/Z = 2}$) in a electron background.

\section{Stellar structure}
\hfill \break

We follow the numerical technique as developed in Refs.~\cite{Bonazzola:1993zz, Bocquet:1995je} to obtain magnetized white dwarf configurations in a fully  general relativity way. Within this approach, the Einstein-Maxwell equations are solved numerically by means of a pseudo-spectra method for axisymmetric stellar configurations within the 3+1 formalism in General Relativity. The accuracy of solutions is controlled by a 2D general-relativistic virial theorem \cite{Bonazzola:1993zz} and, for the stars presented here, it is typically of the order of $10^{-5}$. Recently, this formalism was applied  to neutron stars in Refs.~\cite{franzon2016self, Franzon:2016iai, Franzon:2016urz} and to study magnetized white dwarfs in Ref.~\cite{Franzon:2015gda}. 

With the assumption of a stationary and axisymmetric spacetime, the line element in spherical-like coordinates $(r, \theta, \phi)$ can be written as:
\begin{align}
ds^{2} = &-N^{2} dt^{2} + \Psi^{2} r^{2} \sin^{2}\theta (d\phi - \omega dt)^{2} \\
 &+ \lambda^{2}(dr^{2} + r^{2}d\theta^{2}), \nonumber
\label{line}
\end{align}
with N, $\omega$, $\Psi$ and $\lambda$ being functions of $(r, \theta)$. 
The gravitational field is deduced from the integration of a coupled system of four elliptic partial differential equations for the four metric functions, see e.g. Ref.~\cite{Bonazzola:1993zz}. The final system of gravitational equations can be cast in the form:

\be 
\Delta_{2} [(N \Psi-1) r \sin \theta] = 8\pi N\lambda^2 \Psi r \sin \theta (S^{r}_{r} + S^{\theta}_{\theta}),
\label{Bfinal}
\ee

\be 
\Delta_{2} [{\rm{ln}} \lambda + \nu] = 8\pi \lambda^2 S^{\phi}_{\phi} + \frac{3 \Psi^2 r^2 \sin^2 \theta}{4 N^2} \partial \omega \partial \omega - \partial \nu \partial \nu,
\label{Afinal}
\ee

\be 
\Delta_{3} \nu = 4\pi \lambda^2 (E + S) + \frac{\Psi^2 r^2 \sin^2 \theta}{2N^2} \partial \omega \partial \omega - \partial \nu \partial( \nu + {\rm{ln}} \Psi),
\label{Nfinal}
\ee

and 

\begin{align}
&\left[ \Delta_{3} - \frac{1}{r^2 \sin^2 \theta} \right] (\omega r \sin \theta) \nonumber \\ = & -16\pi \frac{N\lambda^2}{\Psi^2} \frac{J_{\phi}}{r \sin \theta}+ r \sin \theta  \partial \omega \partial(\nu - 3 {\rm{ln}} \Psi),
\label{omegafinal}
\end{align}
where the short notation was introduced:

\begin{align}
& \Delta_{2} = \frac{ \partial^2}{\partial r^2} + \frac{1}{r}\frac{ \partial}{\partial r} + \frac{1}{r^2}\frac{ \partial^2}{\partial \theta^2} \nonumber\\
& \Delta_{3} = \frac{ \partial^2}{\partial r^2} + \frac{2}{r}\frac{ \partial}{\partial r} + \frac{1}{r^2}\frac{ \partial^2}{\partial \theta^2} + \frac{1}{r^2 \tan \theta}\frac{ \partial}{\partial \theta} \nonumber \\
& \nu = {\rm{ln}} N \nonumber.
\end{align}

In addition, in the final gravitational field equations system, Eqs.~\ref{Bfinal}-\ref{omegafinal}, terms as $\partial \omega \partial \omega$ are defined as: 

\be 
\partial \omega \partial \omega := \frac{\partial \omega}{\partial r}\frac{\partial \omega}{\partial r} + \frac{1}{r^2}\frac{\partial \omega}{\partial \theta}\frac{\partial \omega}{\partial \theta} \nonumber, 
\label{notation}
\ee
and the total energy density, momentum density and stress tensors of the system are:
\be 
E = \Gamma^{2} \left( e + p \right) - p + E^{EM}
\ee

\be 
J_{\phi} = \left( E + p \right) \lambda^{2} \Psi  r \sin \theta U + J_{\phi}^{EM}
\ee

\be 
S^{r}_{r} = p + S^{EM\,\,r}_{r} 
\ee 
\be 
 S^{\theta}_{\theta} = p +  S^{EM\,\,\theta}_{\theta}, 
\ee 

\be 
S^{\phi}_{\phi} = p + \left( E + p \right) U^2 + S^{EM\,\,\phi}_{\phi},
\ee
with $e$ and $p$ being the energy and the isotropic pressure of the fluid, $U$ the fluid velocity, $\Gamma$  represents the Lorentz factor which relates the Eulerian and the fluid comoving observers, and $E^{EM}$, $J_{\phi}^{EM}$,  $S^{EM\,\,r}_{r}$, $ S^{EM\,\,\theta}_{\theta}$ and $S^{EM\,\,\phi}_{\phi}$ correspond to the electromagnetic contribution to the energy, momentum and stress tensor of the system, see e.g. Refs.~\cite{franzon2016self,Bocquet:1995je}.

In the case of rigid rotation, the equation of motion ($\nabla_{\mu} T^{\mu\nu}=0$) of a star endowed with magnetic fields
yields:
\be
H \left(r, \theta \right) + \nu \left(r, \theta \right)  + M \left(r, \theta \right) - \Gamma (r, \theta) = const,
\label{equationofmotion}
\ee
where $H(r,\theta)$  is the heat function defined in terms of the baryon number density $n$:
\be
H = \int^{n}_{0}\frac{1}{e(n_{1})+p(n_{1})}\frac{d P}{dn}(n_{1})dn_{1},
\label{heat}
\ee
and the magnetic potential $M(r,\theta)$ can be expressed as:
\be
M \left(r, \theta \right) = M \left( A_{\phi} \left(r, \theta \right) \right): = - \int^{0}_{A_{\phi}\left(r, \theta \right)} f\left(x\right) \mathrm{d}x,
\ee
with $A_{\phi}$ being the magnetic vector potential and $f(x)$ an arbitrary function $f(x)$ that needs to be chosen \cite{Bocquet:1995je}. In our case, we construct stellar models for constant values of $f(x)=k_{0}$. As already shown  by \cite{Bocquet:1995je}, the current function can have a more complex structure, however our general conclusions should remain the same. 

\section{Results}
\hfill \break

In this section, we present the mass-radius diagram for static magnetized white dwarfs by solving self-consistently the Einstein-Maxwell equations, which take into account the anisotropy   of the energy-momentum tensor due to magnetic field.  Recently, studies of modified mass-radius relations of magnetic white dwarfs were proposedin Ref.~\cite{Bera:2014wja}. As we also found in this work, these authors show that the mass of white dwarfs increases in the presence of magnetic fields. 

\begin{figure}[!t]
\center
\includegraphics[width=1.0\textwidth,angle=-90,scale=0.6]{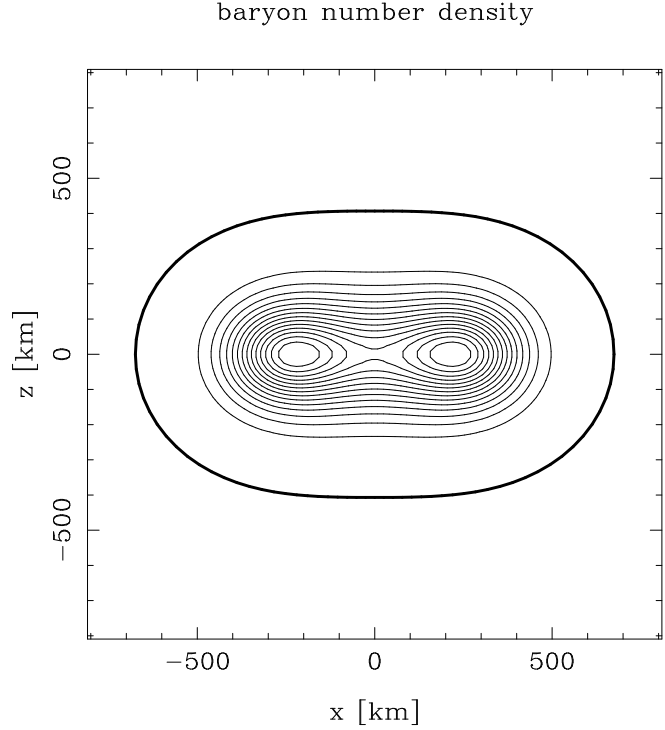}\caption[Isocontours of the baryon number density]{Isocontours of the baryon number density for a star with central magnetic field of $1.03\times 10^{15}$ G.}
\label{nb}
\end{figure}

In Fig.~\ref{nb},  we show the mass density distribution for a static star with central enthalpy  of $\mathsf{H_{c} = 0.0063\,c^{2}}$. This value of the enthalpy results in the maximum gravitational mass of relativistic, static and magnetized white dwarfs achieved within the code, namely,  2.09 $\sf{M_{\odot}}$, which corresponds to a central mass density of $\mathsf{2.79\times 10^{10}\,g/cm^{3}}$. As expected, the mass density is not spherically distributed and the maximum mass density is not at the center of the star. This is due to the fact that  the Lorentz force exerted by the magnetic field  breaks the spherical symmetry of white dwarfs considerably and acts as a centrifugal force that  pushes the matter off-center. In this case, the central magnetic field  reaches a value of $\mathsf{1.03\times 10^{15}}\,$G,  whereas the surface magnetic field was found to be $\mathsf{2.02\times 10^{14}}\,$G. 

\begin{figure}[!t]
\center
\includegraphics[width=1.0\textwidth,angle=-90,scale=0.4]{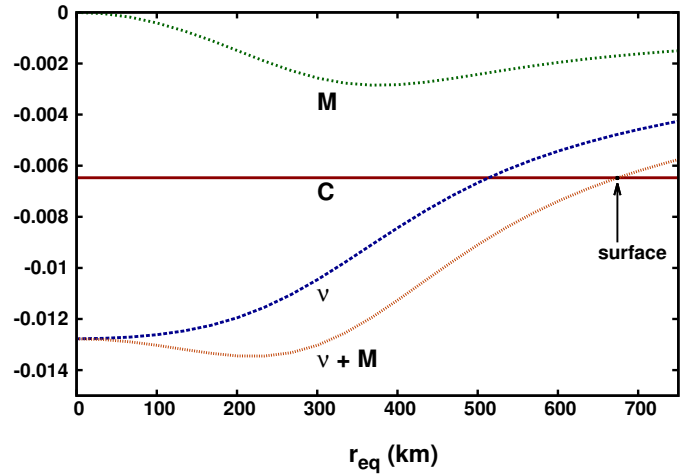}
\caption[Behaviour of the different terms of the equation of motion]{Behaviour of the different terms of the equation of motion as a function of the equatorial coordinate radius for the same star as shown in Fig.~\ref{nb}.}
\label{eqm}
\end{figure}

For a better understanding of this aspect, we make use the equation of motion for the static case, $\rm{\Gamma = 0 }$,
\be
 \mathsf{H(r,\theta) + \nu (r, \theta) +  M(r,\theta) =  C},
\label{bernuli}
\ee
and plot these quantities in the equatorial plane as shown in Fig.~\ref{eqm}. The constant $\mathsf{C}$ can be calculated at every point in the star.  We have chosen the center, since the central value of the magnetic potential $\mathsf{M(r,\theta)}$ is zero and the central enthalpy $\mathsf{H_{c}}$ is our input to construct solutions. The Lorentz force is the derivative of the magnetic potential $\mathsf{M(r, \theta)}$ in the equation Eq.~\eqref{bernuli} and reaches its maximum value off-center ($\mathsf{r_{eq}\sim}$ 350 km, see Fig.~\ref{eqm}).
As already discussed in Ref.~\cite{cardall2001effects}, the  direction of the magnetic forces in the equatorial plane depends on the current distribution inside the star. In addition, the magnetic field changes its direction in the equatorial plane  and, therefore, the Lorenz force reverses the direction inside the star. In our case, this can be seen from  the qualitative change in behaviour of the function $\mathsf{M(r, \theta)}$ around $\sf{r_{eq}} \sim$ 350 km (Fig.~\ref{eqm}). It is worth to mention that we have studied also the effects of the Lorentz force due to high magnetic fields on the geometry of the neutron-star crust in Ref.~\cite{Franzon:2016pbu}, showing that the thickness of the crust can change (either increase or decrease), depending on the polar angle under the influence of the Lorentz force.

\begin{figure}[!t]
\center
\includegraphics[width=1.0\textwidth,angle=-90,scale=0.45]{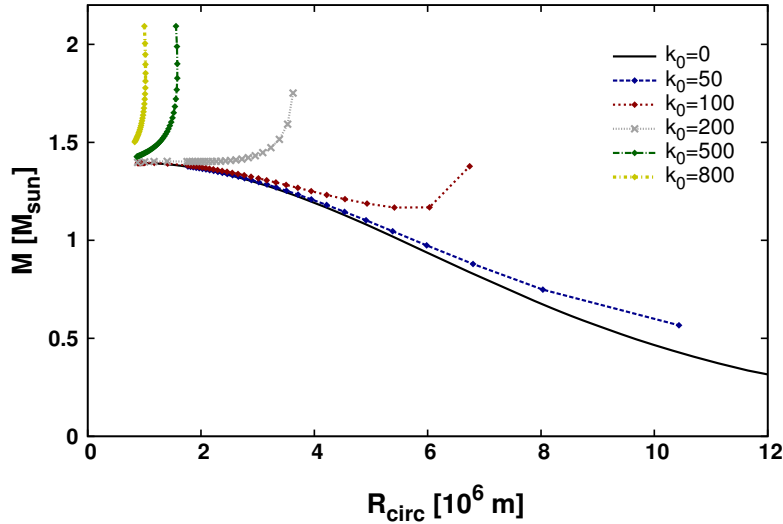}
\caption[Mass-radius diagram for magnetized white dwarfs]{ Mass-radius diagram for magnetized white dwarfs. Different curves represent different values of the current function $\mathsf{k_{0}}$. We also compare the maximum white dwarf mass obtained in the Newtonian case (in black) and in the relativistic one (in red) for the maximum electric current value for which  numerical convergence is achieved. This diagram is quite similar to the MR diagram calculated in \cite{Bera:2014wja}. However, those authors have evaluated only Newtonian white dwarfs. All curves in this figure were calculated for a ratio between the magnetic and matter pressure less than or quite close to 1 at the center of the star.}
\label{wd_static}
\end{figure}

Fig.~\ref{wd_static} depitcs the relation between the mass and the circular equatorial radius for magnetized white dwarfs. The magnetic field is included in the calculation through the current function $\mathsf{k_{0}}$. The higher $\mathsf{k_{0}}$, the higher the magnetic field strength. The stellar sequence (different curves labeled by different $\mathsf{k_{0}'\,s}$) is obtained by changing the central enthalpy of the star (or the central density). The value $\mathsf{k_{0}}=800$  roughly corresponds to the solution with maximum field configuration achieved within the
code. For low magnetic fields, the sequences follow the mass-radius relation as for non-magnetic WD's, ultimately reaching the Chandrasekhar mass limit. As the magnetic field strength is increased, the deviation from the non-magnetic curve increases, resulting in configurations with masses well above the Chandrasekhar limit of 1.4 $\sf{M_{\odot}}$. In this calculation, we found a maximum mass for a relativistic white dwarf of 2.09 $\sf{M_{\odot}}$ for a magnetic field strength at the stellar center of  B $\sim \sf{10^{14}}\,$ G (end of the yellow line in Fig.~\ref{wd_static}).  Fig.~\ref{wd_static} clearly demonstrates that WD masses larger than the standard Chandrasekhar limit for the non-magnetic case can be supported by strong magnetic fields.

Recently, during the writing  process of this thesis, collaborators and I improved  the work presented above, taking into account possible instabilities due to electron capture and
nuclear fusion reactions in the cores of white dwarfs \cite{otoniel2016axisymmetric}. The stellar
interior was composed of a regular crystal lattice made of carbon ions
immersed in a degenerate relativistic electron gas. We found that magnetized white dwarfs
violate the standard Chandrasekhar mass limit significantly,
even when electron capture and pycnonuclear instabilities are present
in the stellar interior. In addition, the maximum magnetic field found is an order of magnitude smaller than in Ref.~\cite{Franzon:2015gda}. This is because we
modeled the stellar interior with a much more realistic equation of
state than just a simple electron gas, and we considered the density threshold for nuclear fusion reactions, which restricts the central density of white dwarfs in $\sim 9.25\times 10^9$ g/cm$^3$, limiting the stellar masses and, therefore,  their radii, which for very massive and magnetized white dwarfs cannot be smaller than $\sim 1100$ km. 

\section{Conclusions}
We were able to calculate super-heavy white dwarfs in the presence of strong fields. This is an interesting and timely important problem, since such stars can exceed the Chandrasekhar mass limit and contribute to superluminous Type Ia supernovae.  
With this in mind, static equilibria of stationary and  axisymmetric white dwarf stars endowed with strong poloidal magnetic fields were carried out. Moreover, we presented a modication on the white dwarf mass-radius relation generated by the magnetic field. We found  that a maximum white dwarf mass of about $\rm{2.00\, M_{\odot}}$ may be supported if the interior  field is as strong as approximately $10^{14}$ G. This mass is over 40 percent larger than the traditional Chandrasekhar limit.

\end{document}